\documentclass[a4paper,11pt]{article}
\usepackage{pos}
\usepackage{float}
\usepackage{arydshln}
\usepackage{colortbl}
\usepackage{amsmath}
\usepackage{dsfont} 
\usepackage{hyperref}
\usepackage{graphicx}
\usepackage{enumitem}
\usepackage{arydshln}
\usepackage{mathtools}
\usepackage{bbold}
\usepackage{soul}
\usepackage{slashed}
\usepackage{amsmath,bm}
\usepackage{yfonts}
\usepackage{lipsum}
\usepackage{slashed}
\usepackage{mathtools}
\usepackage{physics}
\usepackage{tikz}
\usepackage{tikz-feynman}
\usepackage{comment}

\title{Theoretical Aspects of $b \to s \bar{\ell}\ell
$ Decays}
\author*[a]{Arianna Tinari}

\affiliation[a]{Physik-Institut, Universit\"at Zu\"rich, CH-8057 Z\"urich, Switzerland}
\emailAdd{arianna.tinari@physik.uzh.ch}

\abstract{Flavor-changing neutral current decays such as 
$b \to s \bar{\ell} \ell$ are highly suppressed in the Standard Model (SM) 
and therefore provide sensitive tests for new physics. Persistent tensions 
between SM predictions and experimental results in branching ratios and angular observables can be explained by a shift of the Wilson coefficient $C_9$ of the effective operator $\mathcal{O}_9$ by $\sim 20 \%$ relative to the SM value. This shift could arise from a non-standard short-distance contribution or from an inaccurate description of long-distance dynamics, particularly charm rescattering contributions. We therefore investigate charm rescattering contributions in $B^0 \to K^0 \bar{\ell}\ell$ using a model of fundamental hadronic degrees of freedom inspired by
heavy-hadron chiral perturbation theory and improved by appropriate form factors as well as experimental data. Our analysis shows that such effects, with a high degree of fine-tuning, could shift $C_9$ by $\sim 20\%$, at the cost of introducing a more pronounced $q^2$ dependence, whereas experimental data are consistent with a $q^2$-independent shift of $C_9$ with the current experimental uncertainties. In the most natural scenario, we find these effects to be of the order of $\sim 5\%$.
Connections with other flavor anomalies further illustrate the strong discovery 
potential of $b \to s \bar{\ell} \ell$ modes.}

\FullConference{The Thirteenth Annual Large Hadron Collider Physics (LHCP2025)\\
5-9 May 2025\\
Taipei\\}

\begin{document}
\maketitle

\section{Introduction: rare decays}
Rare decays are processes that are forbidden or highly suppressed in the Standard Model (SM). The suppression happens because of loops (for decays that are forbidden at tree-level in the SM), the GIM mechanism, and CKM suppression. Due to this suppression, these decays are highly sensitive to a wide range of new physics (NP) effects, where heavy NP particles can appear at tree- or loop-level in these decays.
One large and interesting class of rare decays is the flavor-changing neutral decays, such as $b\to s, s\to d, c\to u, b\to d$, etc. For these decays, the three suppression factors occur simultaneously. We can study these transitions in $B, D$ or kaon decays. In these proceedings, we focus on the $b \to s \bar{\ell}\ell$ transition.
Here, the observables can be divided into: "clean" observables (lepton-flavor universality ratios), "semi-clean" observables (such as $B_{s, d}\to \bar{\mu}\mu$), and "less clean" observables, such as decay rates and angular distributions.

\section{\texorpdfstring{$b \to s \bar{\ell}\ell$ decays}{}}
\begin{figure}[h]
    \centering
    \includegraphics[scale=0.35]{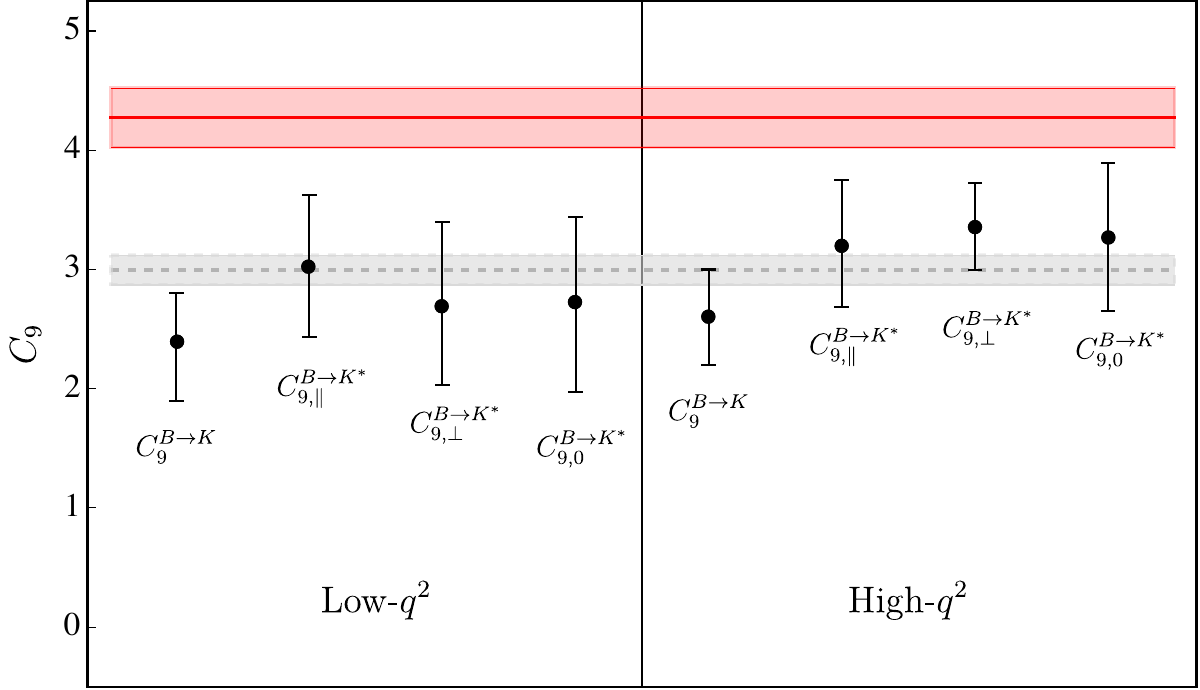}
    \caption{Independent determinations of $C_9$ from data \cite{LHCb:2020lmf, LHCb:2016ykl, CMS:2023klk, LHCb:2014cxe}. The black points illustrate the determinations 
    in the low- and high-$q^2$ regions for the different decay amplitudes.
    The grey band is the result of the fit assuming a universal $C_9$ over the full spectrum. The red 
     band indicates the SM value.  }
    \label{fig:common}
\end{figure}

At the experimental level, there has been a long-standing tension between the SM predictions and the measurements of branching ratios and angular observables of the exclusive modes $B \to K^{(*)}\bar{\ell} \ell, B_s \to \phi \bar{\ell}\ell$. The theoretical predictions are computed within the weak effective theory, valid below the electroweak scale.
The dominant operators relevant for these processes are $\mathcal{O}_7 \propto (\bar{s}_L \sigma_{\mu \nu} b_R) F^{\mu \nu}, \mathcal{O}_9 = (\bar{s}_L \gamma_\mu b_L) (\bar{\ell} \gamma^\mu \ell), \mathcal{O}_{10}=(\bar{s}_L \gamma_\mu b_L) (\bar{\ell} \gamma^\mu \gamma_5 \ell)$. Focusing on $\mathcal{O}_9$, the experimental tension can be explained by a shift in $C_9$ of around 20 \% relative to its SM value. If such a shift comes from short-distance new physics, it should be approximately independent of $q^2$. With the current experimental error, this shift does not show a strong $q^2$ dependence \cite{Bordone:2024hui, Alguero:2023jeh} (see Fig.~\ref{fig:common}). However, this effect could arise from long-distance QCD dynamics, in particular from the matrix elements of the four-quark operators $\mathcal{O}_1 \propto (\bar{s}_L \gamma^\mu T^a c_L)(\bar{c}_L \gamma^\mu T^a b_L) $ and $\mathcal{O}_2 \propto (\bar{s}_L \gamma^\mu c_L)(\bar{c}_L \gamma^\mu b_L)$ that are not well-understood, in which case a $q^2$-dependence is to be expected. 
In particular, charm rescattering effects have been suggested \cite{Ciuchini:2022wbq} as a potentially sizable source of such contributions.

As shown in Fig.~1 in \cite{Ciuchini:2022wbq}, there are two kinds of contributions of charm rescattering; we focus on the second one, the triangle topology. 
Our approach \cite{Isidori:2024lng, Isidori:2025dkp} is to construct a model in terms of hadronic degrees of freedom, based on $SU(3)$ light-quark symmetry and heavy-quark spin symmetry. The goal is to estimate the size of the charm rescattering contributions in the simplest decay mode, $B^0 \to K^0\bar{\ell}\ell$. For this purpose, we employ heavy-hadron chiral perturbation theory combined with QED to describe the kaon and photon vertices, while the $B$-meson couplings are extracted from experimental data on $B$ decays. The resulting description is most reliable at high $q^2$; we then extend it to the full kinematic range by introducing appropriate form factors. Finally, we incorporate the contributions from all possible intermediate states by estimating an overall multiplicity factor.

\section{Model for charm rescattering}
We consider two possible interactions at the photon vertex, the monopole \cite{Isidori:2024lng} and the dipole \cite{Isidori:2025dkp} type. The resulting topologies allowed are shown in Table \ref{tab:diagrams}. The topology associated with the $D^* D^*_s$ intermediate state with the monopole interaction is more complex and requires additional inputs from data, and has not been included in \cite{Isidori:2024lng}. For the dipole coupling, different topologies arise depending on the intermediate states, as shown in Table \ref{tab:diagrams}. The contribution from $B \to D D_s$ vanishes, while those from $B \to D^* D_s^*$ also cancel once all diagrams are summed. The only non-vanishing contributions come from the $B \to D D_s^*$ and $B \to D^* D_s$ channels, yielding four diagrams in total to be evaluated.

\newcolumntype{C}[1]{>{\centering\let\newline\\\arraybackslash\hspace{0pt}}m{#1}}

\begin{table}[t]
    \centering
    \begin{tabular}{C{1.9cm}|C{6.5cm}|C{6.5cm}}
     & monopole interaction & dipole interaction \\
     \hline
     $B \to D D_s$   & not possible &         \begin{tikzpicture}
                \begin{feynman}
                        \node (i1) {$B^0$};
                        \node[right=0.8cm of i1, dot] (i2);
                        \node[right=1cm of i2] (fake);
                        \node[above=0.55cm of fake, dot] (i3);
                        \node[below=0.55cm of fake, dot] (i4);
                        \node[right=0.7cm of i3] (j2) {$K^0$};
                        \node[right=0.7cm of i4] (j3) {$\gamma^*$};
                        \diagram*{
                                (i1) --[scalar] (i2) --[plain] (i4) --[double] (i3) --[plain] (i2) ;
                                (i3) --[scalar] (j2) ;
                                (i4) --[photon] (j3) ;
                        };
                \end{feynman}
        \end{tikzpicture}   \\ \hline
       $B \to D_{s}^* D$, $B \to D^* D_{s}$  &   \begin{tikzpicture}
                \begin{feynman}
                        \node (i1) {$B^0$};
                        \node[right=0.8cm of i1, dot] (i2);
                        \node[right=1cm of i2] (fake);
                        \node[above=0.55cm of fake, dot] (i3);
                        \node[below=0.55cm of fake, dot] (i4);
                        \node[right=0.7cm of i3] (j2) {$K^0$};
                        \node[right=0.7cm of i4] (j3) {$\gamma^*$};
                        \diagram*{
                                (i1) --[scalar] (i2) --[plain] (i4) --[plain] (i3) --[double] (i2) ;
                                (i3) --[scalar] (j2) ;
                                (i4) --[photon] (j3) ;
                        };
                \end{feynman}
        \end{tikzpicture}         \begin{tikzpicture}
                \begin{feynman}
                        \node (i1) {$B^0$};
                        \node[right=0.8cm of i1, dot] (i2);
                        \node[right=1cm of i2] (fake);
                        \node[above=0.55cm of fake, dot] (i3);
                        \node[below=0.55cm of fake, dot] (i4);
                        \node[right=0.7cm of i3] (j2) {$K^0$};
                        \node[right=0.7cm of i4] (j3) {$\gamma^*$};
                        \diagram*{
                                (i1) --[scalar] (i2) --[double] (i4) --[double] (i3) --[plain] (i2) ;
                                (i3) --[scalar] (j2) ;
                                (i4) --[photon] (j3) ;
                        };
                \end{feynman}
        \end{tikzpicture}  & \begin{tikzpicture}
                \begin{feynman}
                        \node (i1) {$B^0$};
                        \node[right=0.8cm of i1, dot] (i2);
                        \node[right=1cm of i2] (fake);
                        \node[above=0.55cm of fake, dot] (i3);
                        \node[below=0.55cm of fake, dot] (i4);
                        \node[right=0.7cm of i3] (j2) {$K^0$};
                        \node[right=0.7cm of i4] (j3) {$\gamma^*$};
                        \diagram*{
                                (i1) --[scalar] (i2) --[double] (i4) --[double] (i3) --[plain] (i2) ;
                                (i3) --[scalar] (j2) ;
                                (i4) --[photon] (j3) ;
                        };
                \end{feynman}
        \end{tikzpicture}       \begin{tikzpicture}
                \begin{feynman}
                        \node (i1) {$B^0$};
                        \node[right=0.8cm of i1, dot] (i2);
                        \node[right=1cm of i2] (fake);
                        \node[above=0.55cm of fake, dot] (i3);
                        \node[below=0.55cm of fake, dot] (i4);
                        \node[right=0.7cm of i3] (j2) {$K^0$};
                        \node[right=0.7cm of i4] (j3) {$\gamma^*$};
                        \diagram*{
                                (i1) --[scalar] (i2) --[plain] (i4) --[double] (i3) --[double] (i2) ;
                                (i3) --[scalar] (j2) ;
                                (i4) --[photon] (j3) ;
                        };
                \end{feynman}
        \end{tikzpicture} \\ \hline
       $B \to D_s^* D^*$   & \begin{tikzpicture}
                \begin{feynman}
                        \node (i1) {$B^0$};
                        \node[right=0.8cm of i1, dot] (i2);
                        \node[right=1cm of i2] (fake);
                        \node[above=0.55cm of fake, dot] (i3);
                        \node[below=0.55cm of fake, dot] (i4);
                        \node[right=0.7cm of i3] (j2) {$K^0$};
                        \node[right=0.7cm of i4] (j3) {$\gamma^*$};
                        \diagram*{
                                (i1) --[scalar] (i2) --[double] (i4) --[double] (i3) --[double] (i2) ;
                                (i3) --[scalar] (j2) ;
                                (i4) --[photon] (j3) ;
                        };
                \end{feynman}
        \end{tikzpicture}        & 
  
       \begin{tikzpicture}
                \begin{feynman}
                        \node (i1) {$B^0$};
                        \node[right=0.8cm of i1, dot] (i2);
                        \node[right=1cm of i2] (fake);
                        \node[above=0.55cm of fake, dot] (i3);
                        \node[below=0.55cm of fake, dot] (i4);
                        \node[right=0.7cm of i3] (j2) {$K^0$};
                        \node[right=0.7cm of i4] (j3) {$\gamma^*$};
                        \diagram*{
                                (i1) --[scalar] (i2) --[double] (i4) --[double] (i3) --[double] (i2) ;
                                (i3) --[scalar] (j2) ;
                                (i4) --[photon] (j3) ;
                        };
                \end{feynman}
        \end{tikzpicture}          \begin{tikzpicture}
                \begin{feynman}
                        \node (i1) {$B^0$};
                        \node[right=0.8cm of i1, dot] (i2);
                        \node[right=1cm of i2] (fake);
                        \node[above=0.55cm of fake, dot] (i3);
                        \node[below=0.55cm of fake, dot] (i4);
                        \node[right=0.7cm of i3] (j2) {$K^0$};
                        \node[right=0.7cm of i4] (j3) {$\gamma^*$};
                        \diagram*{
                                (i1) --[scalar] (i2) --[double] (i4) --[plain] (i3) --[double] (i2) ;
                                (i3) --[scalar] (j2) ;
                                (i4) --[photon] (j3) ;
                        };
                \end{feynman}
        \end{tikzpicture} 
    \end{tabular}
    \caption{Possible topologies for each set of intermediate states for monopole- and dipole-type photon vertex. A single line indicates a $D_{(s)}$, a double line a $D^*_{(s)}$.}
    \label{tab:diagrams}
\end{table}

As mentioned, we also need to introduce some form factors to extend the validity of the calculation to the whole $q^2$ range. These are, in particular, monopole and dipole form factors to extend the point-like QED vertices, and a rescaling of $f_K$ that corrects the $D D^* \to K \bar{\ell}\ell$ amplitude.
The monopole and dipole form factors are extracted from data using a general parametrization that incorporates the tower of charmonium resonances. The free parameters are fixed by fits to $e^+ e^- \to D_{(s)} D_{(s)}^{(*)}$ data \cite{Belle:2006hvs, BESIII:2024zdh}. Additional constraints are imposed at $q^2=0$: in the monopole case, the form factor must reproduce the electric charge, while in the dipole case, the normalization is taken from the lattice QCD calculation of $D_s^* \to D_s \gamma$ \cite{Meng:2024gpd, Donald:2013sra}.

With this setup, we explicitly evaluate the diagrams listed in Table \ref{tab:diagrams}. Their UV divergence is discarded with an $\overline{MS}$-like scheme, and the resulting renormalization scale dependence is used to estimate the uncertainty on the results. 

However, higher resonances with the same $\bar{c}c \bar{s}d$ valence structure are possible. We take these additional states into account by defining a multiplicity factor as follows. For the monopole case, we consider the largest $B \to  X_{\bar{c}c\bar{s}d}$ decays, where $ X_{\bar{c}c\bar{s}d}$ is a possible intermediate state allowed by parity. We then normalize the  $B \to  X_{\bar{c}c\bar{s}d}$ rates to the $B\to D^*D_s+DD^*_s$ one, assuming each of these contributions roughly scales with the corresponding $B \to  X_{\bar{c}c\bar{s}d}$ amplitude with respect to those we
have calculated. We find a multiplicity factor of $\sim 3$ for the monopole case, and by a similar argument based on parity, we find a factor of $\sim 2$ for the dipole case. It should be emphasized that this estimate is deliberately conservative, as we assume that all such contributions add coherently.

\section{Results and conclusions}

\begin{figure} [h]
    \centering
    \includegraphics[width=0.36\linewidth]{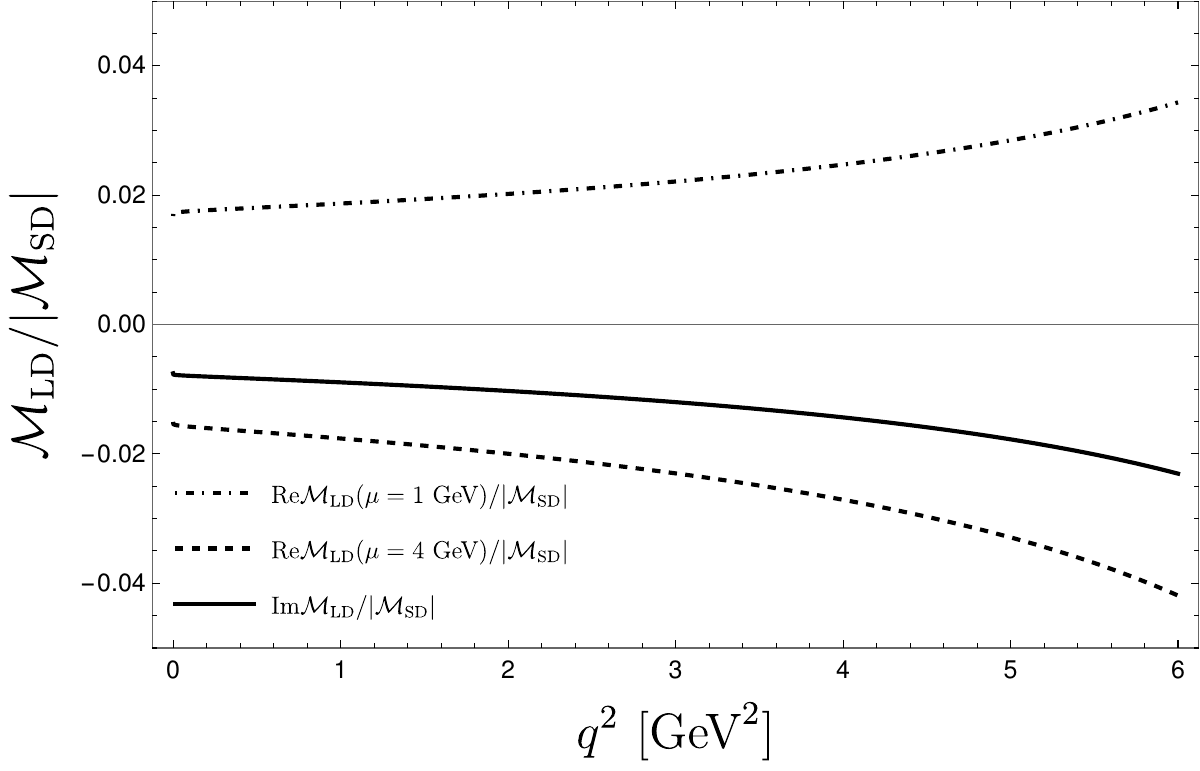}    
    \includegraphics[width=0.36\linewidth]{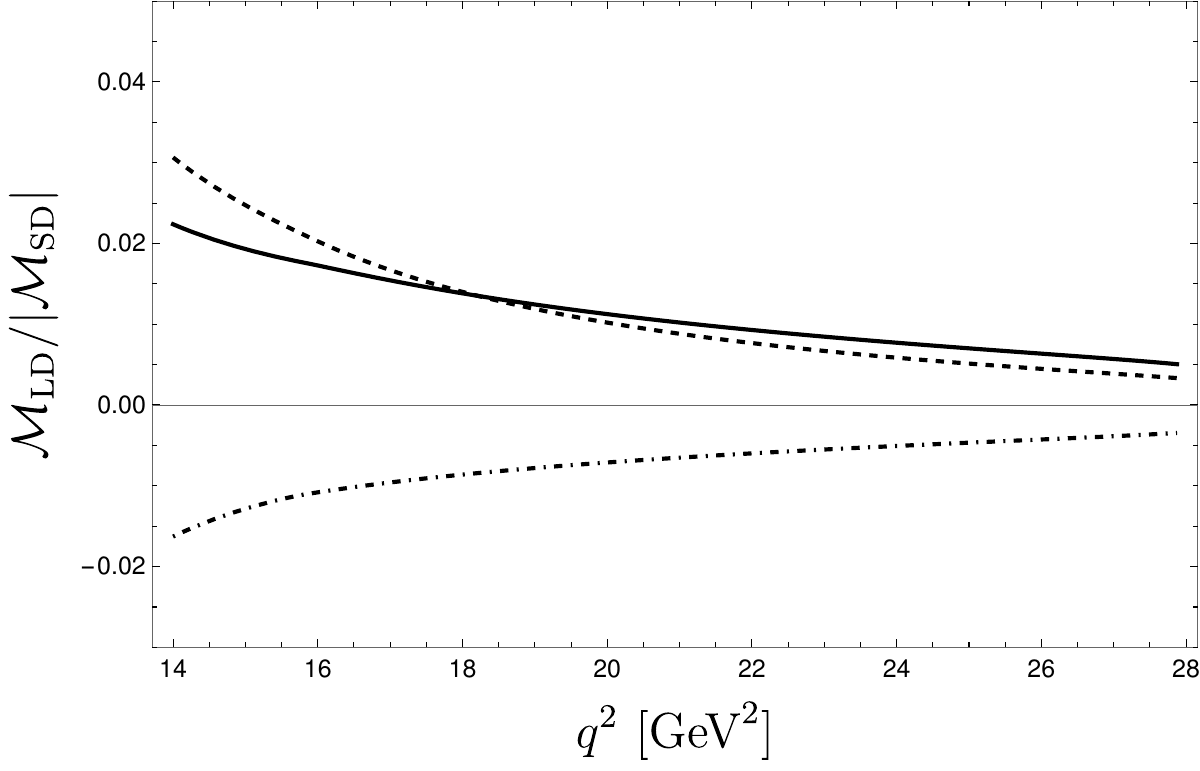}
    \includegraphics[width=0.36\linewidth]{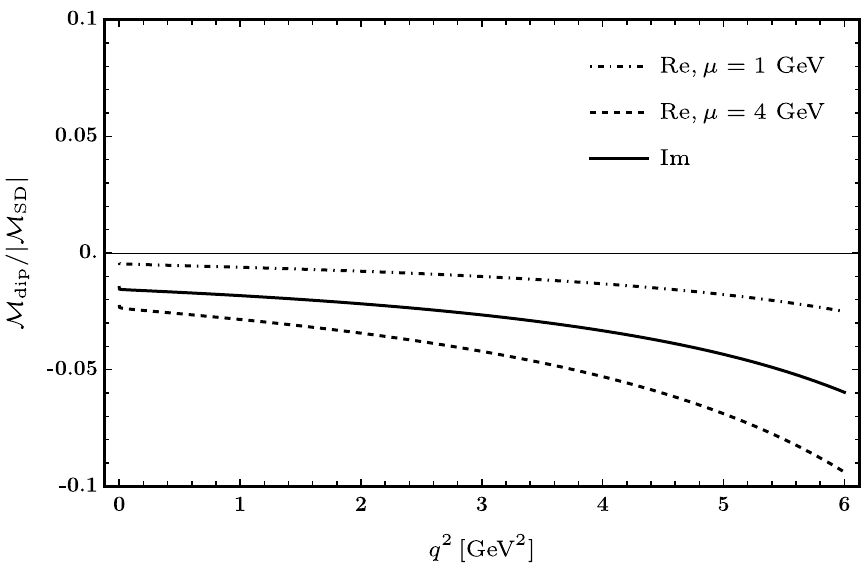}
    \includegraphics[width=0.36\linewidth]{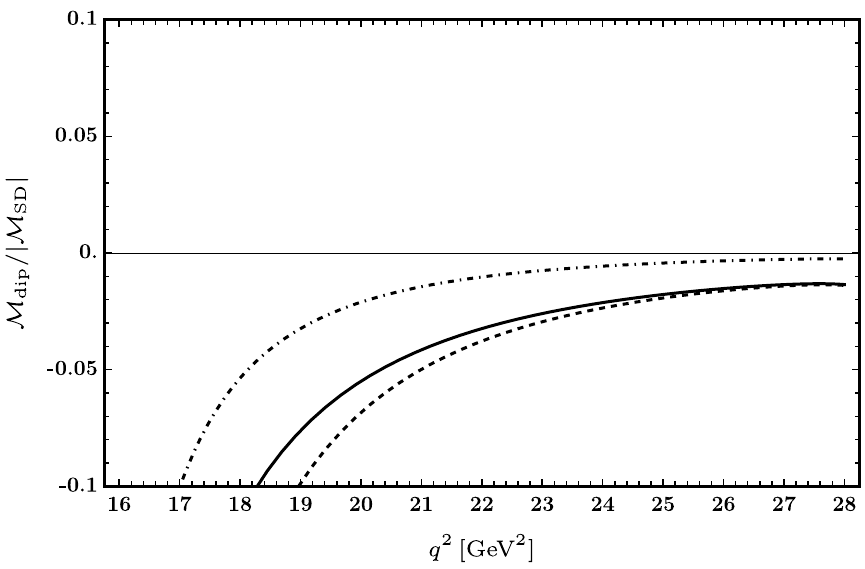}

    \caption{Ratio of the dispersive (dashed and dash-dotted lines) and absorptive (solid line) parts of the monopole (top) and dipole (bottom) matrix element over the absolute value of the short-distance matrix element in the low-$q^2$ (left) and high-$q^2$ (right) regions.}
    \label{fig:ratioMdipMSD}
\end{figure}

The top panels of Fig.~\ref{fig:ratioMdipMSD} show the ratio of the monopole long-distance matrix element to the short-distance matrix element, which is found to be at the level of a few percent. The absorptive part (solid line) is renormalization-scheme independent, as it corresponds to the analytic discontinuity of the amplitude in the kinematic region where the internal mesons can go on shell. For this reason, it can be regarded as model independent. Interpreted as an effective shift in $C_9$, this contribution amounts to about $2$–$3 \%$.
An important feature is that the contribution to $C_9$ changes sign between the low- and high-$q^2$ regions. This sign flip is a generic property of the vector form factor and can be derived under broad theoretical assumptions.
Similarly, the dipole long-distance contributions amount to only a few percent of the short-distance matrix element (see bottom panels of Fig.~\ref{fig:ratioMdipMSD}), though their impact is enhanced in the resonance region.

Since the relative phase between the long-distance contributions and the short-distance one is unknown, we consider three different possibilities for the total contribution to $\delta C_9$:
\begin{enumerate}[label = (\alph*)]
    \item \textbf{Natural:} We assume no tuning, in which case $\delta C_9$ is given just by the absorptive part of the $DD^*$ calculations.
    \item \textbf{Multiplicity-Tuned:} Same as above, but account for additional intermediate states, tuning the relative phases to exactly add coherently.
    \item \textbf{Fully Tuned:} Same as (b), but also assuming maximal interference with the short-distance $C_9$, with a relatively large cutoff scale $\mu/m_D \approx 2$.
\end{enumerate}

\begin{figure*}[h]
    \centering
    \includegraphics[width=0.82\linewidth]{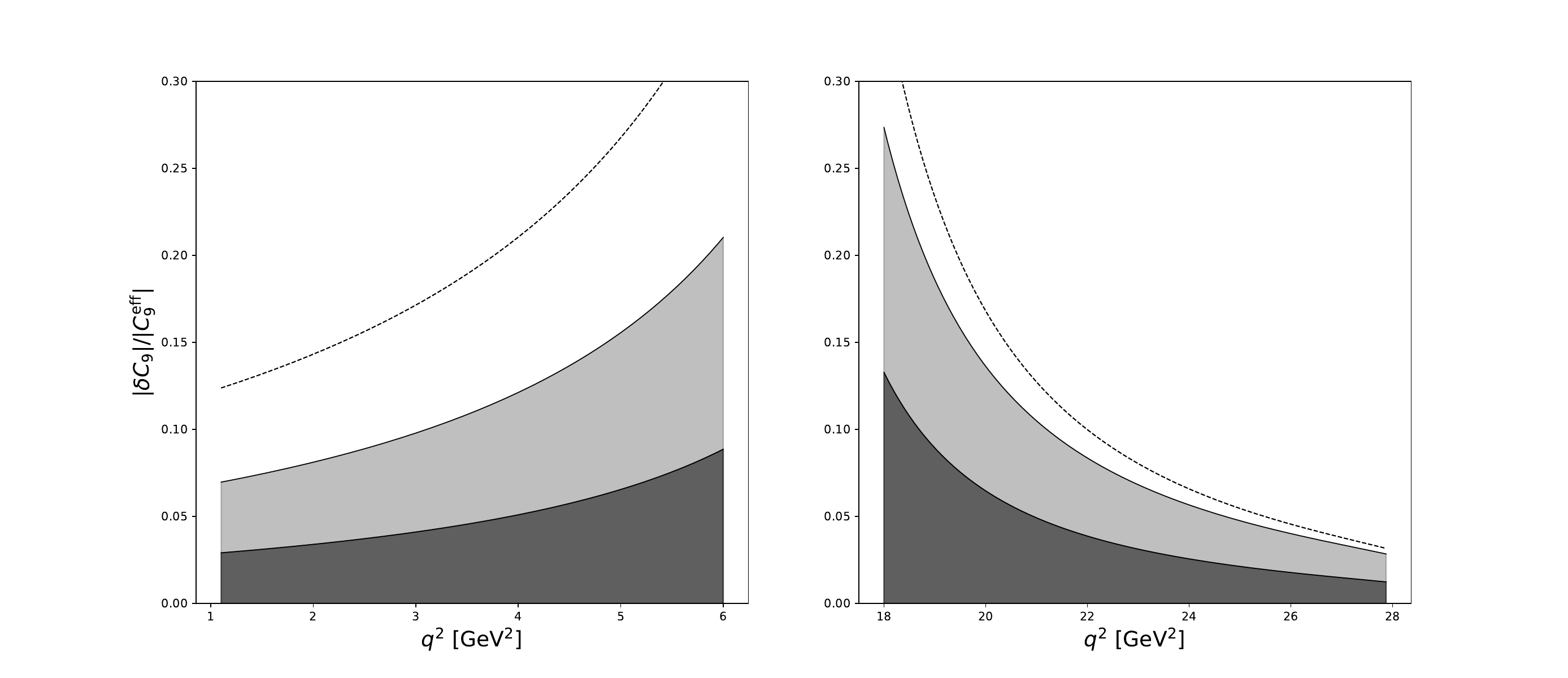}
    \caption{Combined results of monopole and dipole long-distance contributions plotted as $|\delta C_9|/|C_9^{\text{eff}}|$, where $\delta C_9$ is the contribution from charm rescattering triangle diagrams. The dark gray bands give the ``natural'' results, the light gray bands give the partially tuned results, and the dashed lines give the fully tuned results.}
    \label{fig:combinedResults}
\end{figure*}

These three possibilities are plotted in Fig.~\ref{fig:combinedResults}. The results show that it is not unfeasible for rescattering effects to give a sizable, $\sim 20\%$ contribution over a large region of $q^2$, at the cost of a more pronounced $q^2$-dependence, contrary to the hypothesis of Refs.~\cite{Ciuchini:2019usw,Ciuchini:2020gvn,Ciuchini:2022wbq} that unaccounted-for non-local contributions to the matrix elements are mimicking short-distance effects. This is also in tension with the findings of Refs.~\cite{Bordone:2024hui,LHCb:2023gel,LHCb:2023gpo,Alguero:2023jeh} that data are consistent with a $q^2$-independent shift of $C_9$. On the other hand, the ``natural'' case does not exceed an effect of $\sim5\%$ in the region where the modeling of the resonances has the least impact, and in the region near the high-$q^2$ endpoint, where the computation suffers least from model-dependent uncertainties, the contribution to the matrix element does not exceed $\sim5\%$, even in the fully tuned case.

Finally, it is worth mentioning that the $b\to s \bar{\ell}\ell$ anomalies fit into a broader pattern of observables showing tension with the SM, such as $B \to K \bar{\nu} \nu$ \cite{Belle-II:2023esi}, $K \to \pi \bar{\nu}\nu$ \cite{na62}, and $R_{D^{(*)}}$ \cite{RD}. Under the hypothesis of heavy NP, indeed, one can use the framework of SMEFT to explain these tensions. Under the assumption of NP coupling mostly to the third generation of fermions, and an approximate $U(2)$ symmetry acting on the light fermions \cite{Barbieri:2011ci}, only three operators become relevant: $[\mathcal{O}_{lq}^{(1)}]_{3333}=(\bar{l}_3 \gamma^\mu l_3)(\bar{q}_3 \gamma_\mu q_3),[\mathcal{O}_{lq}^{(3)}]_{3333}=(\bar{l}_3 \gamma^\mu \tau^I l_3)(\bar{q}_3 \gamma_\mu \tau^I q_3), [\mathcal{O}_{S}]= (\bar{l}_L^3 \tau_R)( \bar{b}_R q_L^3)$. Adding two parameters that describe the flavor mixing in the quark sector, one can perform a fit with these two free parameters and the Wilson coefficients of the three operators. It turns out that turning on $C_+ \propto C_{lq}^{(1)}+C_{lq}^{(3)}$ with minimal $U(2)$ breaking can explain the di-neutrino modes and is compatible with the $b\to s \bar{\ell}\ell$ bounds \cite{Allwicher:2024ncl}. This illustrates the high discovery potential and discriminating power of the rare modes, especially when analyzed in combination. 

\newpage
\begin{thebibliography}{99}

%\cite{Bordone:2024hui}
\bibitem{Bordone:2024hui}
M.~Bordone, G.~isidori, S.~M\"achler and A.~Tinari,
%``Short- vs. long-distance physics in $B\rightarrow K^{(*)} \ell ^+\ell ^-$: a data-driven analysis,''
Eur. Phys. J. C \textbf{84} (2024) no.5, 547
doi:10.1140/epjc/s10052-024-12869-5
[arXiv:2401.18007 [hep-ph]].
%12 citations counted in INSPIRE as of 23 Sep 2024

%\cite{Alguero:2023jeh}
\bibitem{Alguero:2023jeh}
M.~Alguer{\'o}, A.~Biswas, B.~Capdevila, S.~Descotes-Genon, J.~Matias and M.~Novoa-Brunet,
%``To (b)e or not to (b)e: no electrons at LHCb,''
Eur. Phys. J. C \textbf{83}, no.7, 648 (2023)
doi:10.1140/epjc/s10052-023-11824-0
[arXiv:2304.07330 [hep-ph]].
%119 citations counted in INSPIRE as of 22 Sep 2025

%\cite{LHCb:2020lmf}
\bibitem{LHCb:2020lmf}
R.~Aaij \textit{et al.} [LHCb],
%``Measurement of $CP$-Averaged Observables in the $B^{0}\rightarrow K^{*0}\mu^{+}\mu^{-}$ Decay,''
Phys. Rev. Lett. \textbf{125}, no.1, 011802 (2020)
doi:10.1103/PhysRevLett.125.011802
[arXiv:2003.04831 [hep-ex]].
%479 citations counted in INSPIRE as of 23 Sep 2025

%\cite{LHCb:2016ykl}
\bibitem{LHCb:2016ykl}
R.~Aaij \textit{et al.} [LHCb],
%``Measurements of the S-wave fraction in $B^{0}\rightarrow K^{+}\pi^{-}\mu^{+}\mu^{-}$ decays and the $B^{0}\rightarrow K^{\ast}(892)^{0}\mu^{+}\mu^{-}$ differential branching fraction,''
JHEP \textbf{11}, 047 (2016)
[erratum: JHEP \textbf{04}, 142 (2017)]
doi:10.1007/JHEP11(2016)047
[arXiv:1606.04731 [hep-ex]].
%361 citations counted in INSPIRE as of 23 Sep 2025

%\cite{CMS:2023klk}
\bibitem{CMS:2023klk}
 [CMS],
%``Test of lepton flavor universality in $\mathrm{B}^{\pm}\rightarrow \mathrm{K}^\pm \ell^+ \ell^-$ decays,''
CMS-PAS-BPH-22-005.
%3 citations counted in INSPIRE as of 23 Sep 2025

%\cite{LHCb:2014cxe}
\bibitem{LHCb:2014cxe}
R.~Aaij \textit{et al.} [LHCb],
%``Differential branching fractions and isospin asymmetries of $B \to K^{(*)} \mu^+ \mu^-$ decays,''
JHEP \textbf{06}, 133 (2014)
doi:10.1007/JHEP06(2014)133
[arXiv:1403.8044 [hep-ex]].
%715 citations counted in INSPIRE as of 23 Sep 2025

%%%%%

%\cite{Ciuchini:2022wbq}
\bibitem{Ciuchini:2022wbq}
M.~Ciuchini, M.~Fedele, E.~Franco, A.~Paul, L.~Silvestrini and M.~Valli,
%``Constraints on lepton universality violation from rare B decays,''
Phys. Rev. D \textbf{107}, no.5, 055036 (2023)
doi:10.1103/PhysRevD.107.055036
[arXiv:2212.10516 [hep-ph]].
%124 citations counted in INSPIRE as of 22 Sep 2025

%\cite{Isidori:2024lng}
\bibitem{Isidori:2024lng}
G.~Isidori, Z.~Polonsky and A.~Tinari,
%``Explicit estimate of charm rescattering in B0{\textrightarrow}K0{\ensuremath{\ell}}{\textasciimacron}{\ensuremath{\ell}},''
Phys. Rev. D \textbf{111}, no.9, 093007 (2025)
doi:10.1103/PhysRevD.111.093007
[arXiv:2405.17551 [hep-ph]].
%26 citations counted in INSPIRE as of 19 Sep 2025

%\cite{Isidori:2025dkp}
\bibitem{Isidori:2025dkp}
G.~Isidori, Z.~Polonsky and A.~Tinari,
%``Charm rescattering in $B^0\to K^0\bar{\ell}\ell$: an improved analysis,''
[arXiv:2507.17824 [hep-ph]].
%1 citations counted in INSPIRE as of 19 Sep 2025

%\cite{Belle:2006hvs}
\bibitem{Belle:2006hvs}
K.~Abe \textit{et al.} [Belle],
%``Measurement of the near-threshold e+ e- ---{\ensuremath{>}} D(*)+- D(*)-+ cross section using initial-state radiation,''
Phys. Rev. Lett. \textbf{98}, 092001 (2007)
doi:10.1103/PhysRevLett.98.092001
[arXiv:hep-ex/0608018 [hep-ex]].
%185 citations counted in INSPIRE as of 22 Sep 2025

%\cite{BESIII:2024zdh}
\bibitem{BESIII:2024zdh}
M.~Ablikim \textit{et al.} [BESIII],
%``Precise Measurement of the e+e-{\textrightarrow}Ds+Ds- Cross Section at Center-of-Mass Energies from Threshold to 4.95~GeV,''
Phys. Rev. Lett. \textbf{133}, no.26, 261902 (2024)
doi:10.1103/PhysRevLett.133.261902
[arXiv:2403.14998 [hep-ex]].
%10 citations counted in INSPIRE as of 22 Sep 2025



%\cite{Donald:2013sra}
\bibitem{Donald:2013sra}
G.~C.~Donald, C.~T.~H.~Davies, J.~Koponen and G.~P.~Lepage,
%``Prediction of the $D_s^*$ width from a calculation of its radiative decay in full lattice QCD,''
Phys. Rev. Lett. \textbf{112}, 212002 (2014)
doi:10.1103/PhysRevLett.112.212002
[arXiv:1312.5264 [hep-lat]].
%57 citations counted in INSPIRE as of 22 Sep 2025

%\cite{Meng:2024gpd}
\bibitem{Meng:2024gpd}
Y.~Meng, J.~L.~Dang, C.~Liu, Z.~Liu, T.~Shen, H.~Yan and K.~L.~Zhang,
%``Lattice QCD calculation of the Ds* radiative decay with (2+1)-flavor Wilson-clover ensembles,''
Phys. Rev. D \textbf{109}, no.7, 074511 (2024)
doi:10.1103/PhysRevD.109.074511
[arXiv:2401.13475 [hep-lat]].
%13 citations counted in INSPIRE as of 22 Sep 2025

%\cite{Ciuchini:2020gvn}
\bibitem{Ciuchini:2020gvn}
M.~Ciuchini, M.~Fedele, E.~Franco, A.~Paul, L.~Silvestrini and M.~Valli,
%``Lessons from the $B^{0,+}\to K^{*0,+}\mu^+\mu^-$ angular analyses,''
Phys. Rev. D \textbf{103}, no.1, 015030 (2021)
doi:10.1103/PhysRevD.103.015030
[arXiv:2011.01212 [hep-ph]].
%111 citations counted in INSPIRE as of 22 Sep 2025

%\cite{Ciuchini:2019usw}
\bibitem{Ciuchini:2019usw}
M.~Ciuchini, A.~M.~Coutinho, M.~Fedele, E.~Franco, A.~Paul, L.~Silvestrini and M.~Valli,
%``New Physics in $b \to s \ell^+ \ell^-$ confronts new data on Lepton Universality,''
Eur. Phys. J. C \textbf{79}, no.8, 719 (2019)
doi:10.1140/epjc/s10052-019-7210-9
[arXiv:1903.09632 [hep-ph]].
%218 citations counted in INSPIRE as of 22 Sep 2025

%\cite{LHCb:2023gel}
\bibitem{LHCb:2023gel}
R.~Aaij \textit{et al.} [LHCb],
%``Determination of short- and long-distance contributions in B0{\textrightarrow}K*0{\ensuremath{\mu}}+{\ensuremath{\mu}}- decays,''
Phys. Rev. D \textbf{109}, no.5, 052009 (2024)
doi:10.1103/PhysRevD.109.052009
[arXiv:2312.09102 [hep-ex]].
%36 citations counted in INSPIRE as of 22 Sep 2025


%\cite{LHCb:2023gpo}
\bibitem{LHCb:2023gpo}
R.~Aaij \textit{et al.} [LHCb],
%``Amplitude Analysis of the B0{\textrightarrow}K*0{\ensuremath{\mu}}+{\ensuremath{\mu}}- Decay,''
Phys. Rev. Lett. \textbf{132}, no.13, 131801 (2024)
doi:10.1103/PhysRevLett.132.131801
[arXiv:2312.09115 [hep-ex]].
%37 citations counted in INSPIRE as of 22 Sep 2025

%\cite{Belle-II:2023esi}
\bibitem{Belle-II:2023esi}
I.~Adachi \textit{et al.} [Belle-II],
%``Evidence for B+{\textrightarrow}K+{\ensuremath{\nu}}{\ensuremath{\nu}}{\textasciimacron} decays,''
Phys. Rev. D \textbf{109}, no.11, 112006 (2024)
doi:10.1103/PhysRevD.109.112006
[arXiv:2311.14647 [hep-ex]].
%176 citations counted in INSPIRE as of 23 Sep 2025


\bibitem{na62}
J.~Swallow (NA62)
“New measurement of the $K^+ \to \pi^+\nu^+\nu¯$ decay by the NA62 Experiment, https://indico.cern.ch/event/1447422/





\bibitem{RD}
https://hflav-eos.web.cern.ch/hflav-eos/semi/spring25/html/RDsDsstar/RDRDs.html


%\cite{Barbieri:2011ci}
\bibitem{Barbieri:2011ci}
R.~Barbieri, G.~Isidori, J.~Jones-Perez, P.~Lodone and D.~M.~Straub,
%``$U(2)$ and Minimal Flavour Violation in Supersymmetry,''
Eur. Phys. J. C \textbf{71}, 1725 (2011)
doi:10.1140/epjc/s10052-011-1725-z
[arXiv:1105.2296 [hep-ph]].
%299 citations counted in INSPIRE as of 22 Sep 2025

%\cite{Allwicher:2024ncl}
\bibitem{Allwicher:2024ncl}
L.~Allwicher, M.~Bordone, G.~Isidori, G.~Piazza and A.~Stanzione,
%``Probing third-generation New Physics with K{\textrightarrow}{\ensuremath{\pi}}{\ensuremath{\nu}}{\ensuremath{\nu}}{\textasciimacron} and B{\textrightarrow}K({\textasteriskcentered}){\ensuremath{\nu}}{\ensuremath{\nu}}{\textasciimacron},''
Phys. Lett. B \textbf{861}, 139295 (2025)
doi:10.1016/j.physletb.2025.139295
[arXiv:2410.21444 [hep-ph]].
%24 citations counted in INSPIRE as of 22 Sep 2025



%%%%%%%%%%%%


\end{thebibliography}
\end{document}